\documentclass[a4paper,scriptaddress,twocolumn, prd,showkeys]{revtex4-1}

\usepackage{amssymb} 
\usepackage{amsmath}
\usepackage{amsfonts}
\usepackage{eulervm}

\makeindex

\newcommand{\bra}{\langle}
\newcommand{\ket}{\rangle}

\newcommand{\rd}{\mathrm{d}}

\newcommand{\tr}{\mathrm{Tr}}

\newcommand{\liou}{\mathcal{L}}

\begin{document}

\title{Statistical mechanical expression of entropy production\\ 
for an open quantum system}

\author{Hiroki Majima}
\email{majima@rs.kagu.tus.ac.jp}
\author{Akira Suzuki}

\affiliation
{Department of Physics, Faculty of Science, Tokyo University of Science \\
1-3 Kagurazaka, Shinjyuku-ku, Tokyo 162-8601, Japan}

\begin{abstract}
A quantum statistical expression for the entropy of a nonequilibrium 
system is defined so as to be consistent with Gibbs' relation, 
and is shown to corresponds to dynamical variable by introducing 
analogous to the Heisenberg picture in quantum mechanics. 
The general relation between system-reservoir interactions 
and an entropy change operator in an open quantum system, 
relying just on the framework of statistical mechanics 
and the definition of von Neumann entropy.
By using this formula, we can obtain the correct entropy production 
in the linear response framework. 
The present derivation of entropy production is directly based on the 
first principle of microscopic time-evolution, 
while the previous standard argument is due to the thermodynamic 
energy balance. 
\end{abstract}


\maketitle


\section{Introduction}

The heat generation and entropy production in local 
equilibrium state or nonequilibrium steady state is one of the most 
important and challenging problems in 
nonequilibrium physics.
In most cases 
we are usually interested in the processes of local equilibrium 
and/or nonequilibrium steady state.  
We consider the relevant time scale involved in the processes, 
together with an understanding of the generation of entropy and heat 
in nonequilibrium dissipative systems interacting with an environment 
or coupled quantum systems subject to different time-dependent 
quenches. 
An important physical quantity of the theory of non-equilibrium 
processes is the entropy production that is the rate at which entropy 
is produced as a result of irreversible processes \cite{breuer}.

In the quantum theory of open systems the entropy production 
is usually related to the negative derivative of the relative entropy 
with respect to an invariant state 
in order to show the relative entropy becomes non-negative.
However conventional theories have not succeeded in connecting 
between the entropy production and the interactions 
in open quantum systems. 
Providing a microscopic expression for the entropy production 
has been one of the grand aims of statistical mechanics, 
going back to the seminal work of Boltzmann \cite{boltzmann}. 
A suggestive representation of the nonequilibrium steady-state 
distribution function in terms of the excess entropy production 
has been developed and its connection to the steady-state 
thermodynamics has also been discussed \cite{komatsu}.

In this study, we consider entropy production within the framework 
of linear response theory. 
In statistical physics, Kubo's linear response theory is well known 
as the most effective method, in the linear-approximation regimes, 
of calculating transport coefficients which describe dissipative aspects 
in the microscopic manifestations of microscopic quantum systems. 
For lack of such key-concepts as entropy and/or entropy production, 
however, this theory has long been taken in physics merely as 
a calculational device, without the deep understanding of the reason 
for its general validity.

In our previous work\,\cite{majima1,majima2}, 
we identify certain microscopic interactions inducing heat exchange 
(or friction heat) in dissipative open quantum systems or coupled 
quantum systems. 
By studying the interactions between the system and its environment 
mapped on the representation space , 
we can generally derive a formula 
which connects the interaction Hamiltonian 
to the entropy of the system in open quantum systems 
\cite{majima1,majima2}. 
On the other hand, 
the notion of coarse-grained time-derivative was noticed by Mori 
\cite{mori} and others \cite{kirkwood,ojima}.
Mori especially pointed out that time derivatives appearing in equations 
which define transport coefficients are the 
{\it{average of time derivatives}} 
describing microscopic dynamics \cite{mori}.  
The similar idea of the coarse-graining of time in kinetic 
and transport equations were first given by Kirkwood 
\cite{kirkwood} 
(see e.g., ref.\,\cite{ojima} for a rigorous formulation).

In this paper, we present the relation between the entropy 
production and the Liouvillian superoperator, 
the similar relation was originally derived in the framework of 
thermofield dynamics by the authors \cite{majima1,majima2}, 
and show our relation corresponds to Mori's statistical-mechanical 
expression\, \cite{mori} for the entropy of the nonequilibrium system. 
In the situation like the local equilibrium state or the nonequilibrium 
steady state \cite{leontovich}, where time scale divided into two parts: 
the dynamical variable part and the background part, 
we can define the entropy as a dynamical variable. 
We shall deal with a statistical mechanical expression 
for the entropy of nonequilibrium states, which will be done on the basis 
of quantum statistics. 
The expression for the entropy, which is consistent with Gibbs' relation 
in the thermodynamics of irreversible processes, provides an entropy 
production having the same structure as the entropy production in the 
thermodynamics. 
We also show that the introduction of the entropy as a dynamical variable 
corresponds to definition of physical quantities in Heisenberg picture of 
quantum mechanics. 
In addition, we shall apply the obtained entropy production formula to 
study the effect of interactions within the framework of linear response 
theory. 
As a result, we can obtain a contribution of an interaction Hamiltonian 
to an entropy production rate. 
%
\section{Entropy as a dynamical variable}
%
{\bf{Preliminary:}}  
%
The statistical ensemble of quantum-mechanical systems can be 
described by the density operator $\rho(t)$, which obeys the equation of motion:
\begin{align}
 i\hbar\,\dot{\rho}(t)=[H,\rho(t)] , \label{1}
\end{align}
where $[A,B]=AB-BA$. 
The operator $H$ is the Hamiltonian of the total system and it will be assumed to be a constant 
of motion hereafter. 
The formal integration of Eq. (\ref{1}) is expressed in terms of two time variables $(t,s)$: 
\begin{align}
 \rho(t+s)=U(s)\rho(t)U^{\dag}(s)\,,\,\,\,\, U(s)\equiv\exp(-isH/\hbar)\,.  \label{2}
\end{align}
Therefore, the expectation value of an observable $A$, which does not depend explicitly on the time, 
can then be expressed in the form: 
\begin{align}
\bra A\ket
=\tr\,\rho(t+s)A&=\tr\,\rho(t)A(s)\,, \label{4}
\end{align}
where $\bra\cdots\ket$ indicates the statistical average, and 
$A(s)=U^{\dag}(s)AU(s)$ is the Heisenberg representation 
of the observable $A$, which obeys the equation of motion: 
\begin{align}
 i\hbar\,\dot{A}=[A,H]\,. \label{6}
\end{align}
The expression of the third term in Eq.\,(\ref{4}) is noteworthy for the 
following discussions..

Let us first consider the temporal behavior of observable $A(s)$. 
Here it is important to notice that if the density orator $\rho(t)$ is replaced 
by the local equilibrium density $\rho_t$ defined below, the temporal behavior 
of the observable $A(s)$ starting at a time $t$ for the macroscopic 
state of the system can be discussed by assuming local equilibrium.  
In other words, our distribution which starts from 
the frozen state $\rho_t$ approaches the precise distribution rapidly as 
a result of the microscopic processes with decay time $\tau_0$ due to 
the interaction between mass elements. 
In fact we can consider that the difference vanishes after a lapse of time 
of the order of $\tau_0$. 
This follows from the fact that the difference in the corresponding 
microscopic states is removed by the spontaneous decay of the 
previously-excited microscopic processes. 
In the calculation of the over-all contribution during the time interval 
$\tau$, therefore, the error due to the approximation Eq. (\ref{13}) 
can be estimated to be at most of the order of magnitude of 
$(\tau_0/\tau)|\rho(t)-\rho_t|$ and hence can be neglected if $\tau\gg\tau_0$ 
as in the present case. 
This is the physical basis for the validity of the coarse-graining 
approximation (\ref{13}) \cite{mori2}.  The time derivative 
of the expectation value of an observable $A$ for the time interval $\tau$ 
much shorter than the relaxation time can be defined:
\begin{subequations}
\begin{align}
 \frac{\delta}{\delta t}\bra A(t)\ket
 &\equiv\frac{1}{\tau}\,\bigl[\bra A\ket(t+\tau)-\bra A\ket(t)\bigr] 
 \label{7a}\\ 
 &=\tr\,\,\rho(t)\frac{1}{\tau}\!\int_0^{\tau}\rd s\,U^{\dag}(s)\dot{A}U(s)\,. 
 \label{7c}
\end{align}
\end{subequations}
%

As is well known, the entropy plays an important part in the 
thermodynamical theory of irreversible processes, the starting point 
of which is usually to calculate the irreversible production of entropy 
during the evolution of the irreversible process. 
The relative entropy in particular plays an essential 
role of entropy production in open quantum systems. 
However, the appropriate definition of statistical expression of entropy 
for nonequilibrium states is not known yet in terms of the density 
operator. 
Therefore, in this section, we first define our statistical 
expression of entropy. 

%
\noindent{\bf{Entropy:}}  
If the local equilibrium is assumed at a time $t$, the entropy of the system 
at that time can be given by 
\begin{align}
 S(t)=-k\,\tr\,\,\rho_t\ln\rho_t \label{8}
\end{align}
by virtue of the statistical mechanics of equilibrium states, where $\rho_t$ 
is the local equilibrium distribution. 
The local equilibrium is assumed at discrete values of time whose 
neighboring values are separated by a short time interval $\tau$.  
These local equilibrium points are then connected by the quasi-static 
processes with the help of the procedure of the virtual breaking 
processes \cite{mori}.
Then, Equation (\ref{8}) can be expressed by those discrete values of time 
whose neighboring values are separated by the short time interval $\tau$. 
Accordingly, we can then write $\delta S/\delta t$ in the form:
\begin{align}
 \frac{\delta S}{\delta t}
 =\frac{1}{T}
  \biggl[
   \frac{\delta U}{\delta t}
   -\mu\frac{\delta N}{\delta t}
  \biggr] , \label{9}
\end{align}
where $\delta/\delta t$ means the macroscopic time derivative defined 
in Eq. (\ref{7a}), and $U$, $N$ are respectively the internal energy and the 
particle number. 

We are now in a position to seek a statistical-mechanical expression 
for the entropy of the system at an arbitrary time as is consistent with 
Eqs. (\ref{8}) and (\ref{9}) as follows.

By inserting Eq. (\ref{7c}) into the right-hand side of Eq. (\ref{9}) and 
comparing the obtained equation with an alternative form of the left-hand 
side of Eq. (\ref{9}), we obtain $\delta S/\delta t$ in Eq.\,(\ref{9}) for 
an open quantum system as
\begin{align}
 \frac{\delta S}{\delta t}
 &=k\frac{1}{\tau}\int_0^{\tau}\rd s\,\,\tr\,\,\rho(t) U^{\dag}(s)\frac{1}{kT}
  \Bigl[\dot{H}-\mu\dot{N}\Bigr]U(s)\notag\\
 &=k\,\tr\,\,\rho(t)\frac{1}{\tau}\left[K(\tau)-K(0)\right] \notag\\
 &=k\,\tr\,\,\rho(t)\frac{1}{\tau}\left[U^{\dag}(\tau)KU(\tau)-K\right]\,, 
 \label{10}
\end{align}
where  
\begin{equation}
K\equiv\frac{1}{kT}[H-\mu N]\,.\label{11a}
\end{equation}
From Eq.\,(\ref{10}), we can write $S(t+s)$ in the following form:
\begin{align}
 S(t+s)
 &=S(t)+\tau\frac{\delta S}{\delta t}\notag\\
 &=S(t)+k\,\tr\,\,\rho(t)\left[U^{\dag}(s)KU(s)-K\right]
 \notag\\
 &=S(t)+k\,\tr\,\,\rho(t)K(s)-k\,\tr\,\,\rho(t)K\,, \label{11}
\end{align}
where  the density operator $\rho$ should be replaced by 
\begin{align}
 \rho_t=[\rho(t+s)]_{s=0}=Z^{-1}e^{-K} , \label{13}
\end{align}
for the system in local equilibrium.
Substituting Eq.\,(\ref{10}) in Eq.\,(\ref{11}), we can express 
Eq. (\ref{11}) for the system in local equilibrium as
\begin{align}
 S(t+s)
 &=\tr\,\,\rho_tU^{\dag}(s)[-k\,\ln\rho_t]U(s)\notag\\
 &=\tr\,\,\rho_t\Bigl\{U^{\dag}(s)\bigl[-k\,\ln\rho_t\bigr]U(s)\Bigr\} . 
 \label{14}
\end{align}
Equation (\ref{14}) provides a statistical-mechanical expression 
for the entropy of the nonequilibrium system valid for $t<\tau$, 
which limits the temporal behavior of the entropy to the short time 
interval ($t \sim t+\tau$). 
It is worth to add a remark on the definition of entropy.  
Since our entropy Eq. (\ref{14}) is a dynamical variable like 
\begin{align}
 \bra A\ket (t+s)=\tr\,\, \rho(t) U^{\dag}(s)AU(s) . \label{15}
\end{align}
in contrast to the von Neumann entropy, an entropy operator $\eta$ 
can be defined as 
\begin{align}
 \eta\equiv -k\ln\rho . \label {15.3}
\end{align}

Finally, we point out the relationship between the definition of 
time-dependency of entropy and the pictures in quantum mechanics. 
In the Schr\"odinger picture, entropy $S$ is given by 
\begin{align}
 S(t)=\tr\,\,\rho(t)\eta=-k\,\tr\,\,\rho(t)\ln\rho, \label{15.4}
\end{align}
by introducing the entropy operator Eq. (\ref{15.3}). 
Taking the time derivative of Eq. (\ref{15.4}) , we obtain 
\begin{align}
 \dot{S}(t)=-k\,\tr\,\,\dot{\rho}(t)\ln\rho  \label{15.5}
\end{align}
which corresponds to rate of change with time of the relative entropy 
\begin{align}
 S(\rho(t)|\rho)=-k\,\tr\,\,\rho(t)(\ln\rho(t)-\ln\rho) . \label{15.6}
\end{align}
In the Heisenberg picture, similarly, $S$ is given by 
\begin{align}
 S(t)=\tr\,\,\rho\eta(t)=-k\,\tr\,\,\rho[\ln\rho](t)  \label{15.7}
\end{align}
which is equivalent to our statistical-mechanical expression for the 
entropy Eq. (\ref{14}). 
%
\section{Entropy-Liouvillian relation}
%
In this section we derive the formula which connects the entropy and 
the Hamiltonian of a system by using a definition of entropy operator {\it and} 
the Liouville von Neumann equation. 

In the microscopic picture, 
the state of a system is expressed by a density operator $\rho$. 
For every physical system there exists a Liouvillian superoperator $\liou$, 
which determines the causal evolution of $\rho$ via the following law: 
\begin{align}
 i\hbar\dot{\rho}=\liou\rho\,. \label{16}
\end{align}
Let us introduce an entropy operator as a dynamical variable 
of the system in terms of $\rho$.  
Entropy operator $\eta$, as stated in \S 3, is defined as 
\begin{align}
 \eta=-k\,\ln\rho\,, \label{18}
\end{align}
where the presence of Boltzmann's constant $k$ ensures the Gibbs entropy. 
The expectation value of $\eta$ is then given by
\begin{align}
 \bra\eta\ket=\tr\,\,\rho\eta=-k\,\tr\,\,\rho\ln\rho=S\,. \label{19}
\end{align}
where the statistical average of $\eta$ denoted by a symbol 
$\bra\eta\ket$ is the well-known "Boltzmann-Gibbs entropy" 
if $k$ is taken as the Boltzmann constant.
From Eq.\,(\ref{18}), the density operator $\rho$ may be expressed 
in terms of the entropy operator $\eta$:
\begin{align}
 \rho=e^{-\eta/k}\,, \label{20}
\end{align}
where the time dependency of the density operator $\rho$ 
comes from the entropy operator $\eta$.   
Taking the time derivative of Eq.\,(\ref{20}), we obtain
\begin{align}
 \dot{\rho}=-\frac{\rho}{k} \int_0^1J(\lambda)\rd\lambda\,,\quad
 J(\lambda) \equiv e^{\lambda\eta/k}\dot{\eta}e^{-\lambda\eta/k}\,.
 \label{21}
\end{align} 
Comparing Eq.\,(\ref{16}) with Eq.\,(\ref{21}), we can obtain 
\begin{align}
\rho \int_0^1J(\lambda)\,\rd\lambda 
 =-\frac{k}{i\hbar}\mathcal{L}\rho\,. 
 \label{23}
\end{align}
Taking a trace of Eq.\,(\ref{23}), we obtain
\begin{align}
 \tr
 \biggl[\rho\int_0^1J(\lambda)\rd\lambda\biggr]
 &=-\frac{k}{i\hbar}\tr\,\mathcal{L}\rho\,.\quad\label{24}
\end{align}
Let us expand $K(\lambda)$ with respect to 
$\lambda$: the formal expression is then given by
\begin{align}
 J(\lambda)
 &=\dot{\eta}
   +\lambda\Bigl[\dot{\eta},\frac{\eta}{k}\Bigr]
   +\frac{\lambda^2}{2!}
   \biggl[
    \Bigl[\dot{\eta},\frac{\eta}{k}\Bigr],\frac{\eta}{k}
   \biggr]
   +O(\lambda^3). \label{25}
\end{align}
The second and higher order terms in $\lambda$ arise 
from the fact that the entropy production rate operator $\dot{\eta}$ 
does not generally commute with $\eta$, {\it i.e.}, 
$[\dot{\eta},\eta] \ne 0$, 
which is the characteristic feature of the quantum systems.  
If we retain only the first term in Eq.\,(\ref{25}) 
(note that this approximation is equivalent to the assumption  
$[\dot{\eta},\eta]=0$), Eq.\,(\ref{24}) leads to the entropy 
production of the systems:
\begin{align}
 \bra\dot{\eta}\ket
 =-\frac{k}{i\hbar}\,\tr\,\,\mathcal{L}\rho\,. \label{26}
\end{align}
This relates the {\it entropy production rate} to the {\it Liouvillian} 
$\mathcal{L}$ associated with time evolution of the system and hence 
the entropy production rate for the system can be determined from 
$\tr\,\,\mathcal{L}\rho$.

It should be noted that $[\dot{\eta},\eta] \ne 0$ generally holds 
for the quantum system since $\eta$ is a functional of $\rho(t)$, {\it i.e.}, 
$\eta=\eta[\rho]$.  
The general expression of the {\it entropy-Liouvillian relation} (ELR) 
for the entropy production rate is then given by the following form:
\begin{align}
 \bra\dot{\eta}\ket
 =-\frac{k}{i\hbar}\tr\,\,\mathcal{L}\rho
  +\frac{1}{2k}\bra
    [\eta,\dot{\eta}]\ket
  +O(\lambda^2). \label{27}
\end{align}
The ELR formulas, (\ref{26}) and  (\ref{27}), were systematically 
derived by using the LvN equation (\ref{16}) and the defining equation 
(\ref{18}) for the entropy operator.  
If the density operator $\rho$ is given, we can evaluate the entropy 
production rate $\bra\dot{\eta}\ket$ associated with time-evolution of 
the system. 
%
\section{Entropy production in the linear response scheme}
%
Suppose that a small external field is applied to a system 
of particles at $t=0$. 
The LvN equation is written in the form: 
\begin{align}
 i\hbar\,\dot{\rho}=[H,\rho]. \label{29}
\end{align}
The Hamiltonian, $H$, depends on the time and it consists of 
two parts: $H=H_0+\lambda H_1$, the field-free Hamiltonian $H_0$ with the 
kinetic and potential energy of the particles, and a term $H_1$ for the 
interaction of the particles with the external field. 
Since the external field is weak, the generated change of the density operator 
can be expressed in terms of the expansion in powers of the external field as 
\begin{align}
 \rho=\rho_0+\lambda\rho_1+O(\lambda^2) , \label{30}
\end{align}
Substituting this in (\ref{29}) and comparing each power of 
$\lambda$, we obtain the zeroth-order equation 
\begin{align}
 i\hbar\,\dot{\rho}_0=[H_0,\rho_0]=0.\label{31}
\end{align}
since at $t=0$ a canonical ensemble distribution $\rho_0=Z^{-1}e^{-\beta H_0}$ 
is assumed, where the normalization factor, $Z$, is the partition function.
The corresponding higher-order terms are given by
\begin{align}
 i\hbar\,\dot{\rho}_1=[H_0,\rho_1]+[H_1,\rho_0]\label{32}
\end{align}
and so on. 
The operator $\rho_0$ is the equilibrium distribution of the system 
without the external field. 
The entropy production rate induced by the external field can be 
obtained from the ELR formula (\ref{27}) as follows: 
\begin{align}
 \bra\dot{\eta}_1\ket
 &=-\frac{k}{i\hbar}\tr\,\bigl([H_0,\rho_1]+[H_1,\rho_0]\bigr)
  +O(\eta^2). \label{34}
\end{align}
Since $H_1$ and $H_0$ do not commute, 
the r.h.s of the above equation does not vanish in general.
This means the linear response of the entropy change rate 
is due to  the internal entropy production as well as the external field.
%
%
\bibliographystyle{aipproc}   

\begin{thebibliography}{99}
\bibitem{breuer} 
H. P. Breuer and F. Petruccione, 
{\it The Theory of Open Quantum Systems}, 
(Oxford University Press, Oxford, 2002).
\bibitem{boltzmann} L. Boltzmann. {\it Crelle's Journal}, 
{\bf 98}, 68-94 (1884). 
Reprinted in Hasen\"ohrl (ed.), Wissenschaftlic Abhandlungen, vol. 3. 
New York, Chelsea, pp. 122-152.
\bibitem{komatsu}
T. S. Komatsu and N. Nakagawa, {\it Phys. Rev. Lett.} {\bf 100},  030601 
(2008). 
\bibitem{majima1}
H. Majima, {\it AIP Conf. Proc.} {\bf 985},  756-759 (2008). 
\bibitem{majima2}
H. Majima and A. Suzuki, {\it J. Phys.: Conf. Ser.} {\bf 258}, 012015 
(2010). 
\bibitem{mori}
H. Mori, {\it J. Phys. Soc. Jpn.} {\bf 11}, 1029-1044 (1956). 
\bibitem{kirkwood}
J. G. Kirkwood, {\it J. Chem. Phys.} {\bf 14}, 180-201 (1946). 
\bibitem{ojima}
I. Ojima, H. Hasegawa and M. Ichiyanagi, {\it J. Stat. Phys.} {\bf 56} 
633-655 (1988). 
\bibitem{leontovich} 
M. A. Leontovich, {\it An Introduction to Thermodynamics}, 2nd ed, 
(Gittl Publ, Moscow, 1950 (in Russian)). 
\bibitem{mori2}
H. Mori, {\it Phys. Rev.} {\bf 112}, 1829-1842 (1958). 

\end{thebibliography}


\end{document}